\documentclass[letterpaper,aps,prb,twocolumn,amsmath,showpacs,amssymb]{revtex4}

\usepackage{color}
\usepackage{epsfig}
\usepackage{graphicx}
\usepackage{dcolumn}
\usepackage{bm}
\usepackage{bbm}
\usepackage{wasysym}
\usepackage{marvosym}
\usepackage{times,amsmath,amssymb}

\def\lapp{\lower.35em\hbox{$\stackrel{\textstyle<}{\sim}$}}

\newlength{\textwidthm}

\begin{document}

\title{The infrared conductivity of graphene}

\author{N.~M.~R.~Peres,$^1$ T.~Stauber,$^1$ and A. H. Castro Neto$^2$} 

\affiliation{$^1$Centro de F\'{\i}sica  e  Departamento de
F\'{\i}sica, Universidade do Minho, P-4710-057, Braga, Portugal}

\affiliation{$^2$Department of Physics, Boston University, 590 
Commonwealth Avenue, Boston, MA 02215, USA}

\date{\today}

\begin{abstract}
We study the infrared conductivity of graphene at finite chemical
potential and temperature taking into account the effect of phonons 
and disorder due to charged impurities and unitary scatterers. 
The screening of the long-range Coulomb potential is treated
using the random phase approximation coupled to the coherent
potential approximation. The effect of the electron-phonon coupling is 
studied in second-order perturbation theory. The theory has
essentially one free parameter, namely, the number of
charge impurities per carbon, $n^{{\rm C}}_i$. We find an
anomalous enhancement of the conductivity in a frequency
region that is blocked by Pauli exclusion and an
impurity broadening of the conductivity threshold. We also find that
phonons induce Stokes and anti-Stokes lines that produce 
an excess conductivity, when compared to the far infrared
value of $\sigma_0 = (\pi/2) e^2/h$. 
\end{abstract}


\maketitle


Although there has been enormous experimental \cite{geim_review} and 
theoretical \cite{rmp} progress in understanding the physical properties
of graphene since its isolation in 2004 \cite{Nov04}, the important
issue of graphene transport remains unsettled. More than 20 years ago, 
E. Fradkin \cite{fradkin} showed that the D.C. (static) conductivity of 
graphene, at the charge neutrality point, $\sigma_{{\rm D.C.}}(\mu=0) = 
\sigma(\omega=0,\mu=0)$ ($\omega$ is the frequency and $\mu$ is the
chemical potential measured relative to the Dirac point), cannot be 
described within the standard Boltzmann approach of metals, because the 
Dirac-like electronic excitations have infinite Compton wavelength 
(which is cut-off by the size of the sample), 
violating the assumptions for the
validity of Boltzmann transport \cite{ziman}. 
Fradkin showed that the proper 
way to compute the conductivity is through the Kubo formula treating
the impurities in a self-consistent way. The Kubo formula predicts a
universal, impurity independent, D.C. conductivity: 
$\sigma_{{\rm D.C.,theo.}}(\mu=0) = (4/\pi) e^2/h$. 
Nevertheless, experiments \cite{geim_review} find that 
$\sigma_{{\rm D.C., exp.}}(\mu=0) \approx 4 e^2/h$ with sample-to-sample variations by a factor of 2, which importantly are in the direction of higher conductivity, i.e., further away from the theoretically predicted value
(the so-called mystery of the missing
$\pi$). This result has been assigned to the 
macroscopic inhomogeneity and non-local transport in graphene samples 
\cite{geim_review}. Nevertheless, there is still no consensus in the
theoretical community on the origin of this effect \cite{rmp}. 

\begin{figure}[!ht]
\begin{center}
\includegraphics*[scale=0.4]{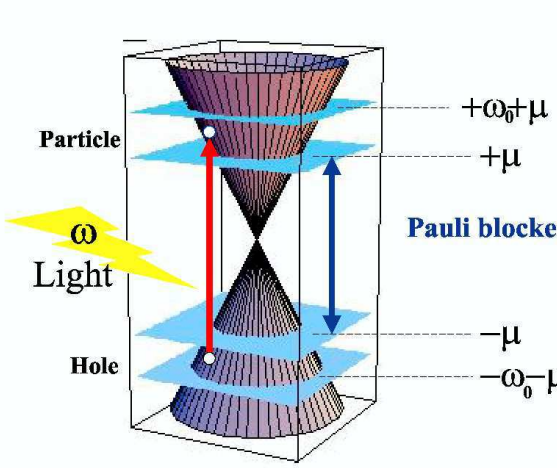}
\end{center}
\caption{ (color online) A photon of frequency $\omega$ creates a
particle-hole pair around the Dirac point. The electronic states with 
energy between $-\mu$
and $+\mu$ are blocked transitions due to the Pauli exclusion principle.
In the figure the phonon frequency $\omega_0$ is assumed to be smaller than
$\mu$. Light with with energy $2 \times (\omega_0 + \mu)$ lead to the
generation of an anti-Stokes emission. The Stokes emission with energy 
$2 \times |\mu-\omega_0|$ lies in the forbidden region. 
\label{cone}}
\end{figure}

In order to settle the issue of transport and scattering mechanisms in
graphene, an aspect of major scientific and technological
significance, it is important to study electronic transport away from
the static regime.  In this regard, the frequency dependent (A.C.)
conductivity, $\sigma(\omega,\mu)$ (we use units such that $\hbar=1$),
provides important information on the scattering mechanisms of the
charge carriers for frequencies $\omega\lapp2\mu$. The basic physical
processes involved in the A.C. conductivity are easy to understand. A
graphene sample is illuminated with light of frequency $\omega$ and
vanishing small wavevector that causes creation of particle-hole pairs
(pair creation), as shown in Fig. \ref{cone}. At zero temperature and
in the absence of disorder or phonons only particle-hole pairs with
energy greater than $2 \mu$ are allowed since all the states with
energy between $-\mu$ and $+\mu$ are forbidden transitions due to
Pauli's exclusion principle. In this case the A.C. conductivity is
simply a step function $\sigma(\omega,\mu) = \sigma_0 \Theta(\omega-
2\mu)$, where $\sigma_0 = (\pi/2) e^2/h$ is the optical conductivity
that has been measured recently \cite{geim_new}.  The far infrared
conductivity is insensitive to phonons, impurities (since these affect
only the low energy part of the spectrum) and band structure effects
when the frequency of incident light is much larger than $2 \mu$ and
much smaller than the electronic bandwidth, $W$ ($\approx 9$ eV)
\cite{nuno_new}.  Nevertheless, as we are going to show, the infrared
spectrum is very sensitive to phonons and impurities and the response
of the system deviates substantially from the non-interacting clean
problem.
  
Phonons of frequency $\omega_0$ can be either absorbed or emitted by the
Dirac electrons. When these phonons are at the center of the Brillouin zone 
($\Gamma$ point) they can be probed by Raman spectroscopy \cite{geim_review},
playing an analogous role as light in an A.C. conductivity experiment,
that is, creation of particle-hole pairs \cite{antonio}. Furthermore,
impurities play a fundamental role at low energies since it is known
that they produce strong broadening of the line-widths \cite{PeresPRB}. 
We stress once more,
as in the case of D.C. transport, that the impurity broadening has to be
calculated self-consistently.
                    
In this paper we compute $\sigma(\omega,\mu)$ taking into account the 
combined effect of impurities and phonons. We assume that there is a 
density $n_i^{{\rm C}}$ of charge
impurities per carbon 
which might be trapped in the substrate, on top of graphene, or in 
the interface of graphene and the substrate.
We model the screening of charge impurities via the random phase 
approximation (RPA) together with the coherent potential approximation (CPA),
which gives us the self-consistent density of states.
We also assume a density $n_i$ of unitary scatterers that exist 
due to structural disorder (edge defects, cracks, vacancies, etc). The effect
of unitary scatterers is only important in producing a finite density
of states at the Dirac point and this can be obtained with arbitrarily small
values of $n_i$. We assume throughout the paper that 
impurities are dilute and the structural disorder is very weak, 
that is, $1 \gg n_i^{{\rm C}} \gg n_i \to 0$. 

We have checked that the effect of 
in-plane acoustic phonons is negligible and they will be ignored in what
follows. We assume that the coupling of graphene to the substrate is
strong enough to shift the flexural phonon frequencies away from the infrared 
regime, allowing us to ignore them for the moment being \cite{note}. Hence,
we have kept only the optical phonon modes. The phonon frequency, and the
value of the electron-phonon coupling is fixed from Raman experiments
\cite{geim_review} and therefore 
they are not fitting parameters here. In fact, 
we have {\it only one fitting parameter}, namely, $n_i^{{\rm C}}$.

The Hamiltonian has the form: 
\begin{eqnarray}
H = H_0 + H_{{\rm ph.}} + H_{{\rm e-ph.}} + H_{{\rm imp.}} \, ,
\label{hamiltonian}
\end{eqnarray}
where
\begin{eqnarray}
H_0=- t \sum_{\bm R,\sigma}\sum_{\bm \delta} 
\left(a^\dag_\sigma(\bm R) b_\sigma(\bm R+\bm \delta)+ {\rm h.c.}\right) \, ,
\label{H0}
\end{eqnarray}
is the nearest-neighbor tight-binding kinetic energy where
$a^\dag_\sigma(\bm R)$ ($b^\dag_\sigma(\bm R+\bm\delta)$) creates
an electron on site ${\bf R}$ of sub-lattice $A$($B$) with spin $\sigma$
($\sigma= \uparrow,\downarrow$), 
$t$ ($\approx 3$ eV) is the hopping energy and 
$\bm\delta$ are the nearest-neighbor vectors \cite{rmp}. 

The phonon Hamiltonian has the form 
\cite{Woods,Suzuura,Ando06,Ishikawa,Neto}:
\begin{eqnarray}
H_{{\rm ph.}} &=& \sum_{{\bf R}} \left\{\frac{{\bf P}_{{\rm A}}^2({\bf
      R})}{2 M_{{\rm C}}} +
\frac{{\bf P}_{{\rm B}}^2({\bf R}+\bm \delta_3)}{2 M_{{\rm C}}} \right.
\nonumber
\\
&+& \sum_{\delta}\frac{\alpha}{2 a^2} \left[({\bm u}_A(\bm
R)-{\bm u}_B(\bm R+\bm\delta))\cdot\bm\delta\right]^2 
\nonumber
\\
&+& \left. \sum_{\delta}\frac{\beta a^2}{2} \left[
\cos(\theta(\bm R,\bm \delta))-\cos(\theta_0)\right]^2 \right\} \, ,
\label{Hph}
\end{eqnarray}
where ${\bm u}_{A,B}$ are the displacements of the A (B) atoms 
from equilibrium (${\bf P}_{A,B}$ the momentum operator), 
$M_{{\rm C}}$ ($=12$ a.u.) is the carbon mass, 
$\alpha$ ($\approx 500$ N/m) is the stretching elastic constant, 
$\theta(\bm R, \bm \delta) = \theta_{ijk}$ is the angle formed between 
the $i-j$ bond and the $i-k$ bond ($\theta_0=120^o$ is the equilibrium angle)
and $\beta$ ($\approx 10 N/m)$ is the in-plane 
bending elastic constant ($a=1.42$ \AA \, is the carbon-carbon distance).
Although (\ref{Hph}) describes both acoustic and optical phonon modes,
we focus on the optical modes which can be written as:
\begin{eqnarray}
{\bm v}({\bf R}) = ({\bm u}_A({\bf R})-{\bf u}_B({\bf R} + \bm
\delta_3))/\sqrt{2} \, ,
\label{optical}
\end{eqnarray}
with frequency $\omega_0^2=3(\alpha+9\beta/2)/M_C \approx 0.2$ eV 
($\approx 1600$ cm$^{-1}$) \cite{Suzuura}.

The electron-phonon Hamiltonian can be written as:
\begin{eqnarray}
H_{{\rm e-ph}} &=&-\frac 1 a \frac {\partial t}{\partial a}
\frac {1}{\sqrt N_c}\sum_{\bm Q,\bm k}\sum_{\sigma,\nu,\bm\delta}
\sqrt{\frac {\hbar}{M_C\omega_\nu(\bm Q)}}
\bm\epsilon_\nu(\bm Q)\cdot\bm\delta
\nonumber\\
&\times&(B^\dag_{-\bm Q,\nu}+B_{\bm Q,\nu})
[e^{i\bm k\cdot\bm\delta}a^\dag_\sigma(\bm k+\bm Q)b_\sigma(\bm k)
\nonumber\\
&+&e^{-i\bm k\cdot\bm\delta}
b^\dag_\sigma(\bm k)a_\sigma(\bm k-\bm Q)]\, ,
\label{Hop}
\end{eqnarray}
where $\partial t/\partial a \approx 6.4$ eV \AA \, is the electron-phonon
coupling, $B^\dag_{\bm Q,\nu}$ creates a phonon of momentum $\bm Q$, 
polarization $\nu$ (polarization vector $\bm\epsilon_\nu(\bm Q)$), 
and frequency $\omega_{\nu}(\bm Q)$ ($N_c$ is the number of unit cells)
\cite{Suzuura}.

The impurity Hamiltonian has the form:
\begin{equation}
H_{imp.}=\frac 1 {N_c}\sum_{\bm p, \bm q,\sigma}V(\bm q)
[a^\dag_\sigma(\bm p)a_\sigma(\bm p+\bm q)+
b^\dag_\sigma(\bm p)b_\sigma(\bm p+\bm q)].
\end{equation}
For a screened impurity of bare charge $Ze$ the potential
$V(\bm q)$ is given \cite{StauberBZ} by 
\begin{eqnarray}
V(\bm q) =
-\frac{Z e^2}{2 \epsilon A_c}\frac{e^{-qd}}{q+\gamma}
\label{Vc}
\end{eqnarray}
where $\epsilon=3.9$ is the SiO$_2$ relative permittivity, $d$ 
is the distance of the impurity to the graphene plane, and $\gamma =
\rho(\mu)e^2 /(2 \epsilon A_c)$ is the RPA screening 
wavevector \cite{rmp} where $\rho(\mu)$ is the self-consistent 
density of states ($A_c=3\sqrt3a^2/2$ is the area of the unit cell). 
Unitary scatterers are modeled using a local potential $V(\bm q) = U$ 
and taking $U \to \infty$. 

\begin{figure}[!ht]
\begin{center}
\includegraphics*[scale=0.3]{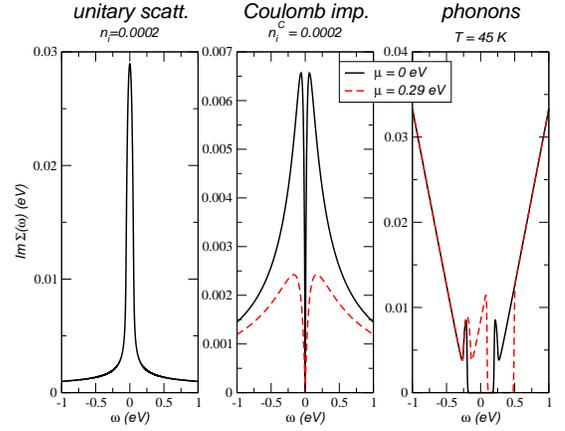}
\end{center}
\caption{
(color online) Imaginary part of the electronic self-energy due 
to unitary scatterers, 
Coulomb impurities, and phonons. The impurity concentration is 
$n_i=n_i^C=2 \times 10^{-4}$. 
We have set $\mu=0$eV (solid line), $\mu=0.29$eV (dashed line) and $T=45$ K.
\label{SE}}
\end{figure}

The effect of a dilute concentration of unitary scatterers can be calculated
exactly using the T-matrix, leading to a
retarded impurity self-energy of the form \cite{PeresPRB}:
\begin{equation}
\Sigma^{{\rm unit.}}_R(\omega) = - \frac{n_i}{
\sum_{\bm k}G_0(\bm k,\omega+i0^+)/N_c} \,,
\label{SCmgs}
\end{equation}  
where $G_0(\bm k,\omega)$ is the free electron Green's function
associated with Hamiltonian (\ref{H0}). Since $\rho_0(\omega) = - 1/\pi 
\sum_{\bm k} \Im G_0(\bm
k,\omega)/N_c \propto  |\omega|$ is the bare
density of states of the clean problem, it is easy to see that (\ref{SCmgs}) 
leads to a divergence of $\Sigma^{{\rm unit.}}_R(\omega \to 0)$ at 
the Dirac point which is unphysical. 
Hence, the problem has to be treated self-consistently
by replacing $G_0(\bm k,\omega)$ by $G^{{\rm unit.}}(\bm k,\omega) 
= G_0(\bm k,\omega)/(1-G_0(\bm k,\omega) \Sigma^{{\rm unit.}}_R(\omega))$ 
in (\ref{SCmgs}). In this case, one can show 
that the self-energy becomes finite at the Dirac point \cite{PLee},
as shown in Fig. \ref{SE}, leading to a finite density of states at zero
energy.

The self-energy due to charged impurities is calculated
in second order perturbation theory as:
\begin{equation}
\Sigma^{{\rm C}}(\bm k,i\omega_n) =\frac { n^C_i}{N_c}
\sum_{\bm p}|V(\bm k-\bm p)|^2G^0(\bm p,i\omega_n)\,,
\label{SEC}
\end{equation}
where a term of the form $n^C_i V(0)$ was absorbed in the definition of
the chemical  potential. The self-energy (\ref{SEC}) is dependent both on the
momentum $\bm k$ and on the frequency. However, we are interested
on the effect of the self-energy for momenta close to the Dirac point
($\bm q = \bm K=\frac{2\pi}{a}(1/3,\sqrt 3/9)$).
Within this approximation, the imaginary part of the
retarded self-energy becomes diagonal
and momentum independent, reading ($d\simeq 0$):
\begin{equation}
\Im\Sigma^{{\rm C}}(\bm K,\omega) \simeq -\frac {Z^2e^4}
{4A_c^2\epsilon_0^2\epsilon}
\frac {n^C_i}{\sqrt 3 t^2} \vert \omega\vert \left(
\frac {2\vert \omega\vert}{3ta}
+\gamma\right)^{-2}\,.
\label{imsig}
\end{equation}
Notice that the imaginary part of the self-energy behaves like
$|\omega|$ at low frequencies and vanishes as $1/|\omega|$ at
large frequencies (see Fig.\ref{SE}). 

Hence, the electron Green's function in the presence of impurities is 
written as: $G^{-1}(\bm k,\omega) = G_0^{-1}(\bm k,\omega) - \Sigma^{{\rm
    unit.}}(\omega) -\Sigma^{{\rm C}}(\bm K,\omega)$. Notice that
the density of states, $\rho(\omega) = - 1/(\pi N_c) \sum_{\bm k} \Im G(\bm
k,\omega)$, should be computed self-consistently since the screening
wavevector $\gamma$ in (\ref{Vc}) and (\ref{imsig}) depends on $\rho(\mu$).

The self-energy due to electron-phonon interaction 
is also computed at the Dirac point in second order perturbation theory:
\begin{align}
\Sigma^{opt}(\bm K,i\omega_n)=
-\frac 9 2 \left(
\frac {\partial t}{\partial a}
\right)^2\frac 1 { M_C\omega_0}
\frac 1 {N_c}\sum_{\bm Q}\notag\\
\times\frac 1{\beta}\sum_m
D^0(\bm Q,i\nu_m)G(\bm K-\bm Q,i\omega_n-i\nu_m)
\end{align}
where $D^0(\bm Q,i\nu_m)= 2\omega_0/((i\nu_m)^2-[\omega_0]^2)$ is the 
phonon Green's function. Notice that $G(\bm k,i\omega_n)$ is the
impurity dressed electronic Green's function. Due to the exclusion principle, 
 the imaginary part of the electron-phonon self-energy
vanishes when $\mu-\omega_0<\omega<\mu+\omega_0$, at $T=0$. At high frequencies
the self-energy follows the electronic density of states and is, therefore, 
linear in $\omega$, as shown in Fig.\ref{SE}.

In the presence of an electromagnetic field the hopping energy
changes to:
\begin{equation}
t\rightarrow te^{ie\bm A(t)\cdot\bm\delta}\,.
\end{equation}
Expanding the exponential up to second order
in the vector potential $\bm A(t)$ and assuming the electric
field to be oriented along the $x$ direction, the current operator is obtained
from $j_x=- \partial H/\partial A_x(t)$ leading to $j_x=j_x^P+A_x(t)j^D_x$.
The Kubo formula for the conductivity is given by: 
\begin{equation}
\sigma(\omega) = \frac{1}{A_s} \frac{1}{i (\omega + i0^+)}
\left[\langle j^D_x\rangle + \Lambda_{xx}(\omega + i0^+) \right] \,,
\end{equation}
with $A_s=N_cA_c$ the area of the sample and 
\begin{eqnarray}
\Lambda_{xx}(i\omega_n) = \int_0^{\beta}d\,\tau e^{i\omega_n\tau}
\langle T_{\tau} j^P_{x}(\tau)j^P_x(0)\rangle \, ,
\end{eqnarray}
 is the current-current correlation function. 
The finite frequency part of the real part of the conductivity is given by:
\begin{equation}
\Re\sigma(\omega)=\frac {2e^2}{\pi h}\int\frac {d\omega'}
{\omega}\Theta(\omega',\omega)[f(\hbar\omega'-\mu)-
f(\hbar\omega'+\hbar\omega-\mu)]\,,
\label{SW}
\end{equation}
where $f(x)$ is the Fermi function and $\Theta(\omega',\omega)$ is a
dimensionless function that depends on the full self-energy and will be 
given elsewhere \cite{PRB}. The main features of the conductivity
can still be understood from Fig.\ref{cone}. Disorder leads to broadening
of the energy levels and a finite density of states at the Dirac point.
This implies that the Pauli exclusion is not effective in blocking
transitions and hence
there is always a finite conductivity even form $\omega < 2 \mu$. The
conductivity in the ``forbidden'' region increases with the increase
in the number of impurities. The fact that the imaginary part of the
electron-phonon self-energy vanishes for electron 
energies between $\mu-\omega_0$ and $\mu+\omega_0$ indicates that for 
$2 \times |\mu-\omega_0| <\omega < 2 \times (\omega_0+\mu)$ the
electron-phonon coupling does not produce any effect in the
conductivity. For $\mu<\omega_0$, we expect the appearance of an
anti-Stokes line at $\omega_{{\rm A.S.}} = 2 \times (\omega_0+\mu)$ and
a Stokes line at $\omega_{{\rm S.}} = 2 \times (\omega_0-\mu)$. 
For $\omega_0<\mu$ the Stokes line lies inside of the Pauli blocked
region and hence it should be suppressed.

\begin{figure}[!ht]
\begin{center}
\includegraphics*[scale=0.3]{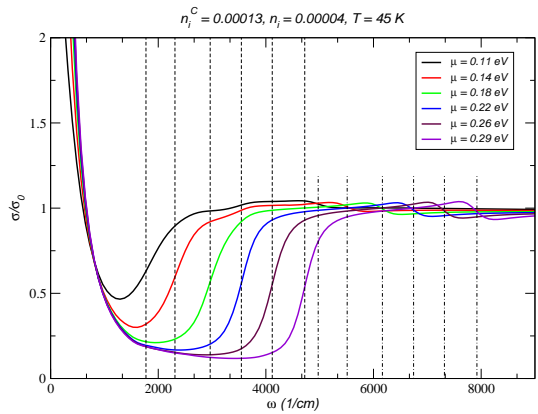}
\end{center}
\caption{(color online)  Real part of the infrared
conductivity including the effect of phonons, unitary scatterers and
charged impurities. The parameters
are $T=45 $ K, $n_i=4.0\times 10^{-5}$, and $n_i^C=1.3\times 10^{-4}$. The
dashed vertical lines correspond to $\omega=2\mu$ and the shorter
dotted-dashed to $\omega=2 (\omega_0+\mu)$ for different values of $\mu$.
\label{Fig_SW}}
\end{figure}

In Figure \ref{Fig_SW} we plot the infrared conductivity of a graphene in
units of the far-infrared conductivity 
$\sigma_0=\pi e^2/(2h)$. The main feature is that the
conductivity is finite in the range $0<\hbar\omega<2\mu$ and increases
as the gate voltage decreases. We choose the concentration of unitary 
scatterers states in Fig. \ref{Fig_SW} to be one order of magnitude 
smaller than
the one of Coulomb scatterers \cite{StauberBZ}, and therefore the
conductivity is mainly controlled by phonons and charged
impurities. Another feature of the curves in Fig. \ref{Fig_SW} is the large
broadening of the inter-band transition edge at $\hbar\omega=2\mu$
(indicated by vertical dashed lines). Note that this broadening is not
due to temperature but to charged impurities, instead.  In fact, the
broadening for all values of $V_g$ is larger when the conductivity is
controlled by charged impurities.
As expected, the coupling to phonons produces a anti-Stokes line centered at
$2 (\omega_0+\mu)$. For gate voltages with
$\mu<\omega_0$ there appears a Stokes line at
$2 (\omega_0-\mu)$. We find, however, that the Stokes line is 
very sensitive to disorder and is fast suppressed by the inclusion
of charge impurities. The optical
phonons thus induce a conductivity larger than $\sigma_0$ around these
frequencies. This effect is washed out at high temperatures and low
frequencies. We also note that for large biases the conductivity in 
the Pauli-blocked region becomes weakly voltage dependent. All these
effects seem to be consistent with the recent infrared measurements
of graphene on a SiO$_2$ substrate \cite{basov}.

In this paper we have studied the infrared conductivity of graphene at
finite chemical potential, generalizing the results of Ref.
[\onlinecite{PeresPRB}]. The calculation includes both the effect of
disorder (unitary scatterers and charged impurities) and the effect of
phonons.  The effect of acoustic phonons is negligible, since it
induces an imaginary part of the electronic self-energy that is much
smaller than the imaginary part induced by either impurities or
optical phonons. We find that optical phonons and charge impurities
produce important modifications in the infrared absorption leading to
large conductivities in the Pauli-blocked energy region of $\omega < 2
\mu$.  The optical phonons also produce a conductivity larger than
$\sigma_0$ around $2\mu< \omega \simeq 2 (\omega_0\pm \mu)$ due to
Stokes and anti-Stokes processes. It is interesting to note that for
frequencies away from the Dirac point the imaginary part of the
self-energy due to optical phonons is linear in frequency, a behavior
similar to that due to electron-electron interactions in graphene
\cite{rmp}. The most important approximation in our calculations is
associated with the fact that we have neglected completely flexural,
modes since we assume that they are pinned by the substrate and hence
have very high excitation energy, that is, away from the infrared
regime.  We stress that the only free parameter in the calculation is
the density of charge impurities, $n^{{\rm C}}_i$ that can change from
sample to sample.

We thank D. N. Basov, A. K. Geim, F. Guinea, P. Kim, and Z. Q. Li for
many illuminating discussions. We thank D. N. Basov, and Z. Q. Li for
showing their data prior to publication.  N.M.R.P. and T.S. were
supported by the ESF Science Program INSTANS 2005-2010, and by FCT
under the grant PTDC/FIS/64404/2006.


\begin{thebibliography}{99}
\bibitem{note}
The effect of flexural modes can be analyzed in a similar way but it
will introduce new free parameters that have to be adjusted experimentally.
Here, we simplify by considering a single adjustable parameter.
\end{thebibliography}
\end{document}